\documentclass[namedreferences]{solarphysics}

\usepackage[hyperref,optionalrh]{spr-sola-addons} 

\usepackage{amsmath,amssymb}
\usepackage{graphicx}
\usepackage{natbib}

\usepackage{xspace}

\usepackage{xcolor}

\usepackage{multirow}

\newcommand{\eg}{{\it e.g.}}

\chardef\us=`\_

\newcommand{\myvect}[1]{{\mathbfit #1}}

\newcommand{\mathi}{\ensuremath{{\rm i}}}

\newcommand{\ptot}{\ensuremath{p_{\rm tot}}}
\newcommand{\ptottilde}{\ensuremath{\tilde{p}_{\rm tot}}}
\newcommand{\Alf}{Alfv$\acute{\rm e}$n}
\newcommand{\Rm}{\ensuremath{R_{\rm m}}}

\newcommand{\xA}{\ensuremath{x_{\rm A}}}

\newcommand{\mi}{\ensuremath{m_{\rm i}}}
\newcommand{\me}{\ensuremath{m_{\rm e}}}

\newcounter{RomanNumber}

\newcommand{\va}{\ensuremath{v_{\rm A}}}
\newcommand{\vai}{\ensuremath{v_{\rm Ai}}}
\newcommand{\vae}{\ensuremath{v_{\rm Ae}}}

\newcommand{\omgR}{\ensuremath{\omega_{\rm R}}}
\newcommand{\omgI}{\ensuremath{\omega_{\rm I}}}

\newcommand{\rhoi}{\ensuremath{\rho_{\rm i}}}
\newcommand{\rhoe}{\ensuremath{\rho_{\rm e}}}

\newcommand{\mathd}{\ensuremath{{\rm d}}}

\begin{document}

\begin{article}
\begin{opening}

\title{Resonant Damping of Kink Modes in Solar Coronal Slabs}

\author[addressref=aff1]{\inits{H.}\fnm{Hui}~\lnm{Yu}}
\author[addressref=aff1,corref,email={bbl@sdu.edu.cn}]{\inits{B.}\fnm{Bo}~\lnm{Li} \orcid{0000-0003-4790-6718}}
\author[addressref=aff1]{\inits{S.X.}\fnm{Shaoxia}~\lnm{Chen}}
\author[addressref=aff1]{\inits{M.Z.}\fnm{Mingzhe}~\lnm{Guo}\orcid{0000-0003-4956-6040}}
\address[id=aff1]{Shandong Provincial Key Laboratory of 
	   Optical Astronomy and Solar-Terrestrial Environment, 
	Institute of Space Sciences, 
	Shandong University, Weihai, 264209 Shandong, China}

\runningauthor{H. Yu et al.}
\runningtitle{Kink Modes in Coronal Slabs}

\begin{abstract}
We examine resonantly damped kink modes in straight coronal slabs, paying special attention to the effects of the formulation for the transverse density distribution (``profile''). We work in the framework of pressure-less, gravity-free, resistive magnetohydrodynamics, and we adopt the dissipative-eigenmode perspective. The density profile is restricted to be one-dimensional, but nonetheless allowed to take a generic form characterized by a continuous transition layer connecting a uniform interior to a uniform exterior. A dispersion relation (DR) is derived in the thin-boundary limit, yielding analytical expressions for the eigenfrequencies that generalize known results in various aspects. We find that the analytical rather than the numerical solutions to the thin-boundary DR serve better the purpose for validating our self-consistent resistive solutions. More importantly, the eigenfrequencies are found to be sensitive to profile specifications, the ratio of the imaginary to the real part readily varying by a factor of two when one profile is used in place of another. Our eigenmode computations are also examined in the context of impulsively excited kink waves, suggesting the importance of resonant absorption for sufficiently oblique components when the spatial scale of the exciter is comparable to the slab half-width.
\end{abstract}

\end{opening}

\section{Introduction} 
\label{sec_intro}

There has been ample observational evidence that the highly structured solar
   corona is replete with low-frequency 
   waves and oscillations
   \citep[see, \eg][for recent reviews]{2012RSPTA.370.3193D,2016SSRv..200...75N,2020ARA&A..58..441N,2020SSRv..216..136L,
   2020SSRv..216..140V,2021SSRv..217...34W}.
A customary practice, known as ``coronal seismology''
   or more broadly ``solar atmospheric seismology'',
   is then to place these observations in the theoretical
   framework of magnetohydrodynamic (MHD) waves in structured media, 
   allowing one to infer the atmospheric parameters that prove
   difficult to directly measure
   (for recent topical collections, see, \eg\
    \citealp{2007SoPh..246....1B};
    \citealp{2009SSRv..149....1N};
	\citealp{2011SSRv..158..167E};
    \citealp{2020FrASS...6...79V}).
From the theoretical perspective, wave-hosting structures have long been 
   modeled as density-enhanced cylinders embedded in an otherwise
   uniform corona
   (\eg\ 
   \citealp{1970A&A.....9..159R};
   \citealp{1975IGAFS..37....3Z};
   \citealp{1982SoPh...75....3S};
   \citealp{1983SoPh...88..179E};
   \citealp{1986SoPh..103..277C};
   see also the recent textbook by 
      \citealp{2019CUP_Roberts}).
The subsequent applications to kink modes in active-region (AR) loops, 
   abundantly measured since the 
   Transition Region and Coronal Explorer era
   \citep[TRACE,][]{1999ApJ...520..880A,1999Sci...285..862N},
   then enabled the inference of the magnetic-field
   strength not only for individual loops \citep{2001A&A...372L..53N}
   but also over a substantial portion of an AR
   \citep{2019ApJ...884L..40A} 
   or even across several ARs \citep{2020Sci...369..694Y}.
While the axial phase speeds alone suffice for this purpose, 
   the damping rates of coronal kink modes have also proved
   informative for inferring the information on the transverse
   density distribution
   \citep[\eg][]{2003ApJ...598.1375A,2007A&A...466.1145A,2008A&A...484..851G,2014A&A...565A..78A},
   provided that the damping can be accounted for
   by the resonant absorption in the \Alf\ continuum
   \citep[\eg][]{2002ApJ...577..475R,2002A&A...394L..39G}. 
This information is known to be critical in wave-based mechanisms for
   coronal heating \citep[\eg][]{2015RSPTA.37340261A,2019ARA&A..57..157C}
   but difficult to glean 
   \citep[see, \eg\ the remarks by][]{2019A&A...622A..44A}. 

There is a long history of theoretical examinations on MHD waves
   in magnetized slabs as well,
   in both non-solar \citep[\eg][]{1973ZPhy..261..203T,1973ZPhy..261..217G,1974PhRvL..32..454H,1974PhFl...17.1399C}
   and solar contexts
   \citep[\eg][]{1978ApJ...226..650I,1979ApJ...227..319W,1982SoPh...76..239E}.
Focusing on the solar context, these examinations are not only important 
    in their own right but prove necessary given the diversity
    of solar atmospheric structures
    \citep[for recent studies, see,  \eg][]{2014A&A...567A..24H,2017SoPh..292...35A,2020ApJ...898...19O}.
Further restricting ourselves to coronal kink modes, 
    we note that a slab configuration is indeed more relevant 
    for interpreting such observations as sunward moving
    tadpoles in post-flare supra-arcades \citep{2005A&A...430L..65V},
    oscillating AR arcades in response to flaring activities
       \citep{2015ApJ...804L..19J,2019ApJ...880....3A},
    and large-scale propagating transverse motions of streamer stalks
       \citep[``streamer waves'',][]{2010ApJ...714..644C,2011ApJ...728..147C,2020ApJ...893...78D}.
The application of a slab equilibrium to streamer waves
    seems somehow surprising given the current sheets embedded in streamer stalks,
    but it is actually justifiable provided that current sheets can be regarded as
    infinitely thin and the electric resistivity can be neglected
    \citep[\eg][]{1986GeoRL..13..373E,2011SoPh..272..119F}.

This study is intended to examine resonantly damped
    kink modes in
    pressure-less, straight coronal slabs
    structured in a one-dimensional (1D) manner, 
    and it can be regarded as an extension of the study by
    \citet[][A07 hereafter]{2007SoPh..246..213A}.
For the ease of discussion, let the slab axis be directed
    in the $z$-direction.
By ``1D'' we refer to a configuration
    where the equilibrium density is nonuniform only
    in the $x$-direction, but it nonetheless takes a rather
    generic form comprising a transition layer
    (TL) that continuously connects a uniform interior
    to a uniform tenuous exterior. 
We further denote the axial
    and out-of-plane wavenumbers    
    by $k_z$ and $k_y$, respectively.    
For the configuration at hand, an extensive list of analytical studies
    has established that
    obliquely propagating ($k_y \ne 0$) kink modes are resonantly
    absorbed in the \Alf\ continuum, 
    provided that the TL width is finite ($l \ne 0$)
    \citep[\eg][]{1978ApJ...226..650I,1979ApJ...227..319W,1988JGR....93.5423H}. 
To our knowledge, these studies tend to assume that
    $k_y \gg k_z$ and $k_y l \ll 1$, which were first lifted
    in the numerical study of A07.
However, among the many observationally relevant parameters, 
    only the influence of $k_y$ was examined therein for a particular
    density profile specification (``profile'' for brevity).
Our study therefore differs from A07 in the following three ways:
First, we will numerically explore a broader set of parameters for the same
    specification and additionally examine two different profiles.
The former aspect is necessary to address, given the evident influence of such
    additional parameters as the TL width on the damping rate. 
On the other hand, although exclusively pertaining to the case where $k_y = 0$,
    previous results have demonstrated the profile sensitivity
    of the dispersive properties of kink modes
    \citep[\eg][]{1988A&A...192..343E,2015ApJ...810...87L,2015ApJ...814...60Y,2018ApJ...855...47C}.
Drawing analogy with pertinent cylindrical studies
    \citep{2013ApJ...777..158S,2014ApJ...781..111S},
    one expects the same sensitivity for a non-vanishing $k_y$
    given that cylindrical results tend to be closely connected to 
    slab ones when $k_y$ is substantial 
    \citep[\eg][]{1992SoPh..138..233G}.
Second, we will also provide analytical expressions for the eigenfrequencies
    in some limiting situations.
While our approach is much-practiced, we will show that these expressions
    extend known solar results by allowing a broader range of $k_y$ 
    and non-solar results by allowing rather general profile choices.
Third, we will connect our computations to impulsively generated kink waves, 
    as likely to be relevant for, say, streamer waves.   
On this aspect we note that resonant absorption has been invoked to account for
    the rapid damping of the oscillatory motions of streamer stalks imaged
    with the COR1 coronagraph on board 
    the Solar TErrestrial RElations Observatory
    \citep[STEREO/COR1,][Figure~6]{2013ApJ...766...55K}.  
We note that the pertinent cylindrical theories were implicated therein,
    despite that a slab configuration seems more appropriate
    \citep[\eg][]{2019ApJ...883..152D}.

This article is organized as follows. 
Section~\ref{sec_model} 
    details the specification of our equilibrium configuration,
    and how we formulate and solve the pertinent eigenvalue problem.
The numerical results are then presented in Section~\ref{sec_num_results}. 
Section~\ref{sec_conc} summarizes the present study, ending 
    with some concluding remarks.

\section{Model Description}
\label{sec_model}

\subsection{Equilibrium Configuration}
\label{sec_equilibrium}
\begin{figure}
\centering
 \includegraphics[width=.9\columnwidth]{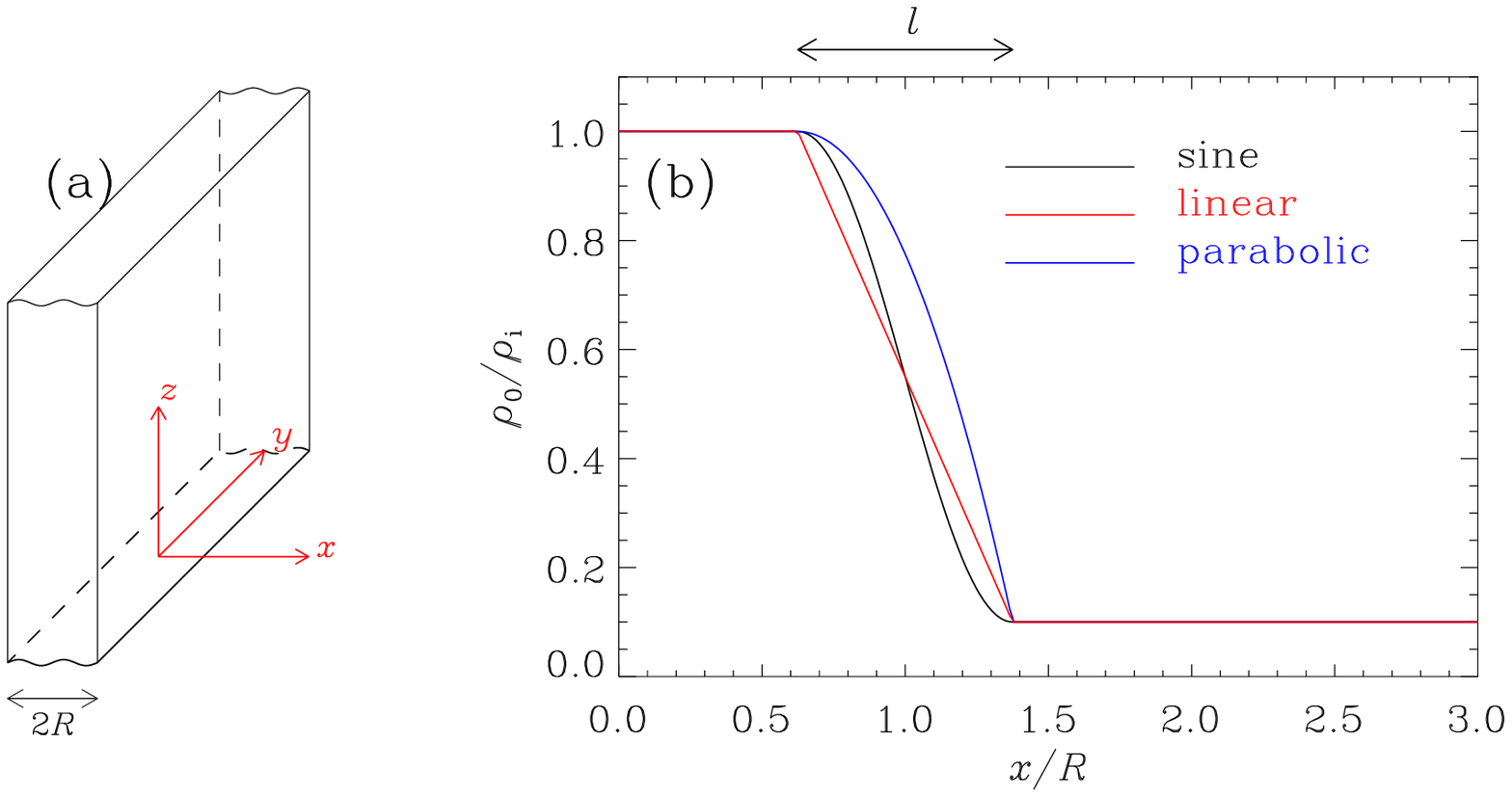}
 \caption{
     (a) Schematic illustration of the equilibrium configuration,
 and (b) transverse distributions of the equilibrium density $\rho_0$.
 All density profiles are characterized
     by a transition layer (TL) that continuously
     connects a uniform interior to a uniform exterior.
 This TL is located between $x_{\rm i} = R-l/2$ and $x_{\rm e} = R+l/2$, where
     $R$ is the mean slab half-width and $l$ the TL width.
 For illustration purposes, 
     the density contrast [$\rhoi/\rhoe$] and the dimensionless TL width [$l/R$]
     are chosen to be $10$ and $0.75$, respectively.
 }
 \label{fig_EQprofile}
\end{figure}

We adopt resistive, gravity-free MHD throughout, 
   and additionally we neglect the gas pressure.
The primitive variables are then the mass density
    [$\rho$], velocity [$\myvect{v}$], and magnetic field [$\myvect{B}$]. 
We denote the equilibrium quantities with the subscript $0$, and assume that
    no equilibrium flow is present ($\myvect{v}_0 = 0$). 
Working in a Cartesian coordinate system $(x, y, z)$, we take the equilibrium magnetic
    field to be uniform and directed in 
    the $z$-direction ($\myvect{B}_0 = B_0 \hat{\mathbf z}$).
We take the equilibrium density [$\rho_0$]     
    to be structured only in the $x$-direction and symmetric about $x=0$.
The transverse profile for $\rho_0$ is further assumed to 
    comprise a uniform interior with density $\rhoi$,
    a uniform exterior with density $\rhoe$,
    and a transition layer (TL) continuously connecting the two
    ($\rhoi > \rhoe$). 
The \Alf\ speed is defined by $v^2_{\rm A} = B_0^2/(\mu_0\rho_0)$, where
    $\mu_0$ is the magnetic permeability in free space. 
We denote the values of $\va$ in the interior and exterior by
    $\vai$ and $\vae$, respectively.     
To specify the density profile,     
    it suffices to consider only the half-space $x \ge 0$,
\begin{equation}
\label{eq_profile_rho_gen}   
\rho_0(x) = \left\{
   \begin{array}{ll}
      \rho_{\rm i},	    	& 0         \le x <   x_{\rm i} = R-l/2, \\ [0.2cm]
      \rho_{\rm tr}(x),		& x_{\rm i} \le x \le x_{\rm e} = R+l/2, \\ [0.2cm]
      \rho_{\rm e},	    	&               x > x_{\rm e}.
   \end{array}
   \right.
\end{equation}
It is evident that this configuration mimics a density-enhanced
   slab of mean half-width $R$ embedded in a uniform
   ambient.
In addition, the TL width [$l$] lies in the range between $0$ and $2~R$.
A number of much-employed TL profiles will be examined, namely
   \citep[\eg][]{2002ApJ...577..475R,2013ApJ...777..158S,2015ApJ...814...60Y}
\begin{equation}
\label{eq_profile_rhoTL} 
\rho_{\rm tr}(x) = 
  \left\{
   \begin{array}{ll}
   \displaystyle
      \frac{\rhoi}{2}
      \left[ \left(1+\frac{\rhoe}{\rhoi}\right)
            -\left(1-\frac{\rhoe}{\rhoi}\right)
             \sin\frac{\pi(x-R)}{l}
      \right],    
                & {\rm sine}, \\[0.3cm]
   \displaystyle 
      \rhoi-\left(\rhoi-\rhoe\right)
            \left(\frac{x-x_{\rm i}}{l}\right),   
                &  {\rm linear},\\[0.3cm]
   \displaystyle                
      \rhoi-\left(\rhoi-\rhoe\right)
            \left(\frac{x-x_{\rm i}}{l}\right)^2,
                & {\rm parabolic}.
   \end{array}
 \right.
\end{equation}
We note that A07 was dedicated to a ``sine'' profile.
Figure~\ref{fig_EQprofile} illustrates both our equilibrium configuration
    and the density profiles.
For illustrative purposes, 
    the density contrast [$\rhoi/\rhoe$] and the dimensionless TL width [$l/R$]
    are chosen to be $10$ and $0.75$, respectively.

\subsection{Formulation of the Eigenvalue Problem and Method of Solution}
\label{sec_linearMHD}

Let the subscript $1$ denote small-amplitude perturbations to the equilibrium.     
The linearized resistive MHD equations then read 
\begin{align}
\label{eq_linMHD_momen}
  \displaystyle
   \rho_0 \frac{\partial \myvect{v}_{1}}{\partial t}
     &= \frac{(\nabla\times\myvect{B}_1)\times \myvect{B}_0}{\mu_0},
                    \\ 
\label{eq_linMHD_Farad}                     
  \displaystyle
   \frac{\partial \myvect{B}_1}{\partial t}
     &= \nabla\times\left(\myvect{v}_1\times \myvect{B}_0-\frac{\eta}{\mu_0}\nabla\times\myvect{B}_1
                   \right). 
\end{align}
Here $\eta$ is the Ohmic resistivity, taken to be constant for simplicity.
We adopt an eigenvalue-problem (EVP) standpoint by  
    Fourier-analyzing any perturbation as
\begin{equation}
\label{eq_Fourier}
   f_1 (x, y, z; t)
 = {\rm Re}\{\tilde{f}(x)\exp[-\mathi (\omega t- k_y y -k_z z)]\},
\end{equation}
   where $\omega$ is the angular frequency, and $k_z$ ($k_y$)
   represents the axial (out-of-plane) wavenumber.
With tilde we denote the Fourier amplitude.         
We take $k_y$ and $k_z$ as real-valued, but allow $\omega$ to be complex-valued. 
If some quantity is complex, then we denote
    its real and imaginary parts with subscripts~R and I, respectively.
Note that $\omgI < 0$ throughout.    
In component form, Equations~\ref{eq_linMHD_momen}
	and \ref{eq_linMHD_Farad} become
\begin{align}
\label{eq_Fourier_vx}
\displaystyle
\omega\tilde{v}_x & =  
     -\frac{B_0}{\mu_0 \rho_0}
      \left(k_z \tilde{B}_x + \mathi \tilde{B}'_z\right), \\ 
\label{eq_Fourier_vy}
\displaystyle                 
\omega\tilde{v}_y & =  
     -\frac{B_0}{\mu_0\rho_0}
      \left(k_z \tilde{B}_y - k_y \tilde{B}_z\right), \\
\label{eq_Fourier_Bx}       
\displaystyle				 
\omega\tilde{B}_x & =  
     -B_0 k_z \tilde{v}_x 
     +\frac{\mathi\eta}{\mu_0}
        \left[\tilde{B}_x''-\left(k_y^2+k_z^2\right)\tilde{B}_x
        \right], \\
\label{eq_Fourier_By}                 		  
\displaystyle				 
\omega\tilde{B}_y & =  
     -B_0 k_z \tilde{v}_y 
     +\frac{\mathi\eta}{\mu_0}
        \left[\tilde{B}_y''-\left(k_y^2+k_z^2\right)\tilde{B}_y
        \right], \\
\label{eq_Fourier_Bz} 
\displaystyle				 
\omega\tilde{B}_z & =  
     -\mathi B_0 \left(\tilde{v}_x'+ \mathi k_y \tilde{v}_y\right) 
     +\frac{\mathi\eta}{\mu_0}
        \left[\tilde{B}_z''-\left(k_y^2+k_z^2\right)\tilde{B}_z
        \right],         		  
\end{align}
    where we have used the shorthand notation
    $' \equiv \mathd/\mathd x$.
Equations~\ref{eq_Fourier_vx} to \ref{eq_Fourier_Bz}
    constitute a standard EVP when supplemented with
    appropriate boundary conditions (BCs).
With kink modes in mind, we specify the BCs
    at the slab axis ($x=0$) as
     $\tilde{v}_x'    = \tilde{v}_y 
      = \tilde{B}_x'  = \tilde{B}_y = \tilde{B}_z = 0$.
Only trapped modes are of interest, in accordance with which 
    all variables are required to vanish when $x \to \infty$.  

\begin{figure}
\centering
 \includegraphics[width=.9\columnwidth]{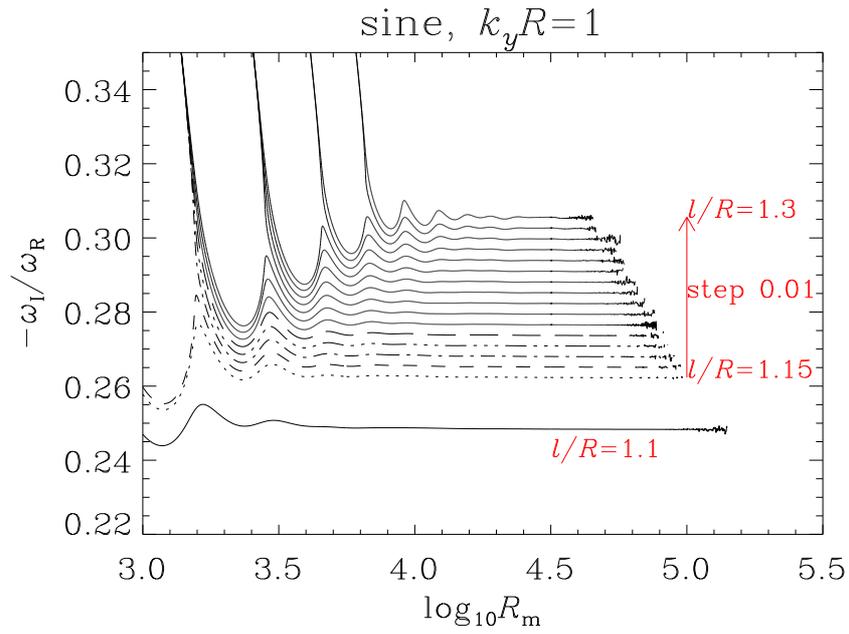}
 \caption{
 Dependence on the magnetic Reynolds number [$\Rm$] 
     of the ratio of the imaginary to the real part of the eigenfrequency
     [$-\omgI/\omgR$] of resonantly damped kink modes in coronal slabs.
 A ``sine'' profile is adopted for the equilibrium density,
     and the density contrast [$\rhoi/\rhoe$] is chosen to be $10$.
 A number of values are examined for the dimensionless transition
     layer width [$l/R$], with $l/R$ evenly spaced 
     by $0.01$ with the exception of the curve labeled $l/R=1.1$.         
 The axial and out-of-plane wavenumbers correspond to 
     $k_z R = \pi/50$ and $k_y R = 1$, respectively.           
 }
 \label{fig_omg_Rm_solproc}
\end{figure}

The solution procedure for the EVP is as follows.
We start with normalizing Equations~\ref{eq_Fourier_vx}
    to \ref{eq_Fourier_Bz}, for which purpose 
    we take $R$, $\vai$, and $\rhoi$ as
    the independent normalizing constants. 
The derivative normalizing constants for time and magnetic field
    are chosen to be $R/\vai$ and $B_{\rm i} \equiv \sqrt{\mu_0 \rhoi \vai^2}$,
    respectively. 
The Ohmic resistivity is wrapped up in
    the magnetic Reynolds number $\Rm = \mu_0 R \vai/\eta$.             
We formulate and solve the EVP
    with the general-purpose finite-element code PDE2D~\citep{1988Sewell_PDE2D},
    which was first introduced into the solar context
    by \citet{2006ApJ...642..533T} to our knowledge.
We adopt a computational domain of $[0, x_{\rm M}]$, 
    and place the outer boundary [$x_{\rm M}$] sufficiently far from the slab
    such that further increasing $x_{\rm M}$ does not influence
    our numerical results.
A nonuniform grid is employed to save computational cost, 
    and a considerable fraction of the grid points is deployed 
    in the TL to resolve the possible oscillatory behavior therein.
It turns out that the details of the grid setup have some bearings on 
    the parameter range that we can explore.
To illustrate this, we recall that the resistive approach here is not intended to
    explore the full spectrum of resisitive eigenmodes.
Rather, we are interested only in the one that falls back to the standard kink mode
    when the TL is infinitely thin ($l=0$) and the resistivity vanishes
    ($\eta = 0$).
Allowing $l$ to be finite, we are adopting the well known resistive eigenmode
    approach to examine the essentially ideal process of the resonant absorption
    of a collective mode in the relevant continuum
    \citep[\eg][]{1991PhRvL..66.2871P,2004ApJ...606.1223V,2006ApJ...642..533T,2016SoPh..291..877G,2018ApJ...868....5C}
	or continuaa \citep[\eg][]{2009ApJ...695L.166S,2021ApJ...908..230C}.
As such, we are looking for only those eigenfrequencies that do not depend on $\Rm$
    when $\Rm$ is sufficiently large.
This is illustrated by Figure~\ref{fig_omg_Rm_solproc}
    where the ratio of the imaginary to the real part of the eigenfrequency
    (or $-\omgI/\omgR$ to be precise) 
    is shown as a function of the magnetic Reynolds number [$\Rm$]
    for a ``sine" density profile with $\rhoi/\rhoe = 10$.
In addition, $[k_y R, k_z R]$ is fixed at $[1, \pi/50]$.
A series of $l/R$, equally spaced by $0.01$, is examined except for
    the lowermost curve.
The outer boundary is fixed at $x_{\rm M} = 50~R$.
Furthermore, $18000$ uniformly spaced grid points are employed for $x\le 3~R$,
    beyond which the grid spacing increases by a constant factor of $1.00125$. 
For a given $l/R$, one sees that the eigenfrequencies are indeed $\Rm$-independent
    over an interval of $\Rm$, for which the 
    left and right ends are to be denoted by $R_{\rm m, L}$ and $R_{\rm m, R}$,
    respectively.
Evidently, the sought-after eigenfrequencies can be confidently identified only
    when the interval $[R_{\rm m, L}, R_{\rm m, R}]$ is sufficiently broad.
However, this cannot be  guaranteed. 
On the one hand, $R_{\rm m, L}$ tends to increase with $l/R$.     
On the other hand, our code does not converge when $\Rm$ becomes too large
   for the well known reason that the eigenfunctions become increasingly oscillatory
   in the so-named dissipative layers (DLs)
   (see, \eg\ Figure~4 in 
   \citealp{1995JPlPh..54..129R};
   Figure~1 in 
   \citealp{1996ApJ...471..501T}).
On top of that, Figure~\ref{fig_omg_Rm_solproc} indicates that some zigzags appear
   when $\Rm$ approaches the maximal value that the code can handle, the end result
   being that the interval $[R_{\rm m, L}, R_{\rm m, R}]$ 
   narrows with $l/R$.
Overall, we find that this numerical difficulty arises 
   when the damping is heavy, which occurs when $l/R$ and/or $k_y R$ are 
   large.
Numerically speaking, this interval can be made broader by employing
   more grid points to resolve the fine scales in the DLs.
In practice, we experiment with a substantial number of
   grid setups such that we can
   expand the parameter range examined by A07 where the ``sine'' profile is adopted.
In addition, we will examine another two profiles
   stated in Equation~\ref{eq_profile_rhoTL}.
The purpose of presenting Figure~\ref{fig_omg_Rm_solproc}
   is to illustrate the $\Rm$-dependence of the eigen-frequencies
   of resonantly damped kink modes in coronal slabs,
   thereby complementing cylindrical studies with similar scopes
   (\eg, Figure~2 in 
          \citealp{2006ApJ...642..533T}; 
          Figure~9 in 
          \citealp{2016SoPh..291..877G}).

For ease of description, we formally express the eigenfrequencies as
\begin{equation}
\label{eq_omega_formal}
\displaystyle 
  \frac{\omega R}{v_{\rm Ai}} 
= {\cal G}
  \left(
       {\rm prof}, \frac{\rhoi}{\rhoe}, \frac{l}{R};
       k_y R, k_z R 
  \right),
\end{equation}
    where ``prof'' represents a prescription for the equilibrium density profile.   
The magnetic Reynolds number does not appear in Equation~\ref{eq_omega_formal},
    which means in practice that the computed eigenfrequencies are found
    in an interval of $[R_{\rm m, L}, R_{\rm m, R}]$ spanning 
    at least half a decade or so.

\subsection{Resonantly Damped Kink Modes in the Thin-Boundary Limit}
\label{sec_TB}
The functional dependence in Equation~\ref{eq_omega_formal}
    can be established analytically in a number of situations.
This subsection is intended to expand some available results, for which purpose
    we start by noting that the ideal version ($\eta = 0$)
    of Equations~\ref{eq_Fourier_vx} to \ref{eq_Fourier_Bz} can be
    manipulated to yield a single equation governing $\tilde{v}_x$
    (\eg\ A07), namely
\begin{equation}
\label{eq_2nd_order_vx}  
\displaystyle 
	\left[
	   \frac{k_z^2 - \omega^2/v_{\rm A}^2}{k_y^2 + k_z^2 - \omega^2/v_{\rm A}^2}
	   \tilde{v}_x'
	\right]'
  - \left(k_z^2 - \omega^2/v_{\rm A}^2\right)\tilde{v}_x
  = 0.
\end{equation}
Evidently, the \Alf\ resonance takes place at 
    $x=\xA$ where $\omgR = k_z \va$.
We now distinguish between two schools of approaches
    for treating resonantly damped modes.
In one school, broadly referred to as the ideal quasi-mode approach,
    dissipative effects are mathematically irrelevant and 
    a non-vanishing $\omgI$ is accounted for by, say, analytically continuing
    the Green function 
    (for early studies in various contexts, see, \eg\
    \citealp{1973ZPhy..261..203T};
    \citealp{1973ZPhy..261..217G};
    \citealp{1974PhRvL..32..454H};
    \citealp{1974PhFl...17.1399C};
    \citealp{1978ApJ...226..650I};
    \citealp{1979ApJ...233..756W};
    see also the review by
    \citealp{2011SSRv..158..289G}).
Equation~\ref{eq_2nd_order_vx} is involved in a substantial number of 
    quasi-mode studies, which tend to 
    assume $k_y^2 \gg k_z^2$ and $k_y^2 \gg  |k_z^2 - \omega^2/v_{\rm A}^2|$ from the outset, and address a ``linear'' profile.
Of particular relevance is the one in the fusion context
    by \citet[][hereafter TW98; see also references therein]{1998JPSJ...67.2322T}, 
    who showed that adopting a linear profile enables one
    to derive an analytical dispersion relation (DR) valid for arbitrary TL widths.
As a result, the resonant damping is allowed to be heavy.     
Another school,
    referred to as the dissipative eigenmode approach 
    and hence involving dissipative factors by construction,
    does not restrict the range of $k_y$ but nonetheless assumes weak damping 
    by working in the so-called thin-boundary (TB) limit ($l/R \ll 1$).
However, no restriction is necessary for the equilibrium profile    
    \citep[\eg][to name only a few]{1991SoPh..133..227S,1992SoPh..138..233G,1995JPlPh..54..129R}.     
We will follow the second approach.

We proceed by defining
\begin{align}
\label{eq_def_kappaie}
\displaystyle 
&  \kappa^2_{\rm i, e} 
= k_z^2 - \frac{\omega^2}{v^2_{\rm Ai, e}},  \\
\label{eq_def_mie}
\displaystyle 
&  m^2_{\rm i, e} 
= k_y^2 + k_z^2 - \frac{\omega^2}{v^2_{\rm Ai, e}}, 
\end{align}
   where 
   $-\pi/2 < \arg \kappa_{\rm i, e},~\arg m_{\rm i, e} \le \pi/2$.
Trapped kink modes then correspond to the following solution to
   Equation~\ref{eq_2nd_order_vx} in the uniform portions   
\begin{equation}
\label{eq_sol_vx_uniform_regions} 
\tilde{v}_x(x) = 
  \left\{
   \begin{array}{ll}
   \displaystyle
      A_1 \cosh(\mi x),    
                & 0 < x < x_{\rm i}, \\[0.3cm]
   \displaystyle 
      A_2 \exp(-\me x),
                & x > x_{\rm e},
   \end{array}
 \right.
\end{equation}    
    where $A_1$ and $A_2$ are constants. 
In addition, the Fourier amplitude of the Eulerian perturbation of total 
    pressure ($\ptot = \myvect{B}_0\cdot\myvect{B}_1/\mu_0 = B_0 B_{1z}/\mu_0$)
    reads
\begin{equation}
\label{eq_sol_ptot_uniform_regions} 
\ptottilde(x) = 
  \left\{
   \begin{array}{ll}
   \displaystyle
      \left(\frac{B_0^2}{\mu_0}\right)
      \left(\frac{-\mathi A_1}{\omega}\right) 
      \left(\frac{\kappa_{\rm i}^2}{\mi}\right)
      \sinh(\mi x),    
                & 0 < x < x_{\rm i}, \\[0.3cm]
   \displaystyle 
      \left(\frac{B_0^2}{\mu_0}\right)
      \left(\frac{\mathi A_2}{\omega}\right) 
      \left(\frac{\kappa_{\rm e}^2}{\me}\right)
      \exp(-\me x),    
                & x > x_{\rm e}.
   \end{array}
 \right.
\end{equation} 
Let $\{q\}$ denote the variation of some perturbation $q$ across some thin 
    dissipative layer bracketing the resonance. 
Furthermore, let $\tilde{\xi}_x$ denote the Fourier amplitude of the $x$-component
    of the Lagrangian displacement ($\tilde{v}_x = - \mathi \omega \tilde{\xi}_x$).
Assuming $l/R \ll 1$, the TB approximation attributes the variation of $q$ 
    across the TL to $\{q\}$, and 
    $\{\ptottilde\}$ and $\{\tilde{\xi}_x\}$ were found to be
    \citep[\eg][]{1996ApJ...471..501T,2000ApJ...531..561A}
\begin{align}
\label{eq_jump_ptot}
& \{\ptottilde\} = 0,~ 			 \\
\label{eq_jump_xi}
& \{\tilde{\xi}_x\} = 
     -\frac{\mathi\pi k_y^2 {\rm sgn}(\omgR)}{\rho_{\rm A} |\Delta_{\rm A}|}
      \tilde{p}_{\rm tot, A},~
\end{align}     
   where any quantity designated with a subscript A is evaluated
   at the \Alf\ resonance.
Equations~\ref{eq_jump_ptot} and \ref{eq_jump_xi}
   are a slab generalization of the cylindrical version established
   by \citet{1991SoPh..133..227S}.
The symbol $\Delta_{\rm A}$ was also introduced therein, and reads for the
   present slab configuration
\begin{equation}
\Delta_{\rm A} =
  \left.
       \frac{{\rm d}(\omega^2-k_z^2 v_{\rm A}^2)}{{\rm d}x}\right|_{x=x_{\rm A}}.
\end{equation}   
With the jump conditions (Equations \ref{eq_jump_ptot} and \ref{eq_jump_xi}), 
   one can connect $\tilde{v}_x$ and $\ptottilde$ in the interior
   and exterior,
   thereby establishing the following DR
\begin{equation}
\label{eq_DR_TB}
\displaystyle 
   \coth(\mi R) 
= -\left(\frac{\kappa_{\rm i}^2}{\kappa_{\rm e}^2}\right) 
   \left(\frac{\me}{\mi}\right) 
  +\mathi\pi 
   \left(\frac{k_y^2}{k_z^2}\right)
   \left(\frac{\rho_{\rm A}}{\rho'_{\rm A}}\right) 
   \left(\frac{\kappa_{\rm i}^2}{\mi}\right).   			
\end{equation}
Here $\omgR$ is taken to be positive without loss of generality.
We have also used the fact that $\va^2(x) \propto 1/\rho(x)$ 
   and $\rho' < 0$  in our equilibrium setup.

Equation~\ref{eq_DR_TB} is analytically tractable 
   when one assumes that $k_y^2 \gg |\omega^2/\va^2|$, in which case
   $\mi \approx \me \approx k_y$. 
Further assuming that $k_y l \ll 1$ and $|\omgI| \ll \omgR$, one proceeds 
   by keeping only the terms linear in $\omgI$.
The real part of the right-hand side (RHS) of
   Equation~\ref{eq_DR_TB} is dominated by the first term, resulting in
\begin{equation}
\label{eq_TB_omgR}
\displaystyle 
   \omgR^2 
 = k_z^2 \vai^2
   \frac{1+\Theta}{\rhoe/\rhoi+\Theta},   		
\end{equation}    
   where $\Theta = \tanh(k_y R)$. 
With the aid of Equation~\ref{eq_TB_omgR}, 
   the leading terms of the imaginary part of  
   Equation~\ref{eq_DR_TB} lead to that
\begin{equation}
\label{eq_TB_omgI_tmp}
\displaystyle 
   \frac{\omgI}{\omgR}
= -\frac{\pi}{2}
   \left(k_y \left|\frac{\rho_{\rm A}}{\rho'_{\rm A}}\right|\right)
   \frac{(1-\rhoe/\rhoi)^2\Theta^2}{(\rhoe/\rhoi+\Theta)^2 (1+\Theta)}.   		
\end{equation}
As intuitively expected, $\omgR$ as given by Equation~\ref{eq_TB_omgR}
   does not involve the specific prescription 
   for the density profile.
Consequently, the density at the resonance always evaluates to 
   $\rho_{\rm A} = (\rhoe + \Theta\rhoi)/(1+\Theta)$.   
However, the profile specification does influence $\omgI$ through the
    $\rho'_{\rm A}$-term in Equation~\ref{eq_TB_omgI_tmp}.
To evaluate this term, it can be readily found that the resonance takes place at
\begin{equation}
\label{eq_TB_xA} 
\displaystyle 
\frac{\xA - R}{l} = 
  \left\{
   \begin{array}{ll}
   \displaystyle
   \frac{1}{\pi}\arcsin\left(\frac{1-\Theta}{1+\Theta}\right), 
                & {\rm sine}, \\[0.3cm]
   \displaystyle
   \frac{1}{1+\Theta}-\frac{1}{2},
                & {\rm linear}, \\[0.3cm]
   \displaystyle 
   \left(\frac{1}{1+\Theta}\right)^{1/2}-\frac{1}{2},   
                & {\rm parabolic}.
   \end{array}
 \right.
\end{equation}
It then follows that
\begin{equation}
\label{eq_TB_rhoprime_tmp1}
\rho'_{\rm A} = -\left(\frac{\rhoi-\rhoe}{l}\right) \frac{1}{g},
\end{equation}
   where
\begin{equation}
\label{eq_TB_g} 
\displaystyle 
  g 
= 
  \left\{
   \begin{array}{ll}
   \displaystyle
   \frac{1+\Theta}{\pi\sqrt{\Theta}}, 
                & {\rm sine}, \\[0.3cm]
   \displaystyle
   1,
                & {\rm linear}, \\[0.3cm]
   \displaystyle 
   \left(\frac{1+\Theta}{4}\right)^{1/2},   
                & {\rm parabolic}.
   \end{array}
 \right.
\end{equation}
Eventually, the TB damping rate can be expressed as
\begin{equation}
\label{eq_TB_omgI}
\displaystyle 
  \frac{\omgI}{\omgR}
= -\frac{\pi}{2}
   \left(k_y l\right) g
   \frac{(1-\rhoe/\rhoi) \Theta^2}{(\rhoe/\rhoi+\Theta) (1+\Theta)^2}.
\end{equation}

Some remarks are necessary to address the assumption $k_y l \ll 1$.
When deriving Equation~\ref{eq_TB_omgR}, we neglected 
   the second term on the RHS of Equation~\ref{eq_DR_TB}.
Now with Equation~\ref{eq_TB_omgI_tmp} available, we can 
   go back to Equation~\ref{eq_DR_TB}
   and evaluate the real part of the second term on its RHS.
To maintain consistency, this real part should be far smaller than the LHS, 
   which evaluates to $1/\Theta$.
The end result is that the following inequality should be valid,
\begin{equation*}
   \displaystyle 
   \left[\pi k_y \left(\frac{\rho_{\rm A}}{\rho'_{\rm A}}\right)\right]^2 
   \frac{(1-\rhoe/\rhoi)^2 \Theta^3}{(\rhoe/\rhoi+\Theta)^3} \ll 1.   
\end{equation*}       
Specializing to the profiles in this study, 
   we find with Equation~\ref{eq_TB_rhoprime_tmp1} that
\begin{equation*}
   \displaystyle 
   \left(k_y l\right)^2 
   \left(\pi g\right)^2 
   \frac{\Theta^3}{(1+\Theta)^2 (\rhoe/\rhoi+\Theta)} \ll 1.
\end{equation*}
By requiring $k_y l \ll 1$ we actually mean the situations where
   the afore-mentioned inequalities hold, which may be less restrictive.

We now compare our TB expressions with some previous results.
We start by considering a further limit where
   $k_y R$ is large enough to ensure $\Theta \approx 1$.
In this case, Equation~\ref{eq_TB_omgR} simplifies to
\begin{equation}
   \label{eq_TB_omgk}
\omgR^2 = \omega_{\rm k}^2 \equiv k_z^2 c_{\rm k}^2~,
\end{equation}
   where
\begin{equation}
    \label{eq_def_ck}
\displaystyle 
c_{\rm k}^2 = \frac{2 \vai^2}{1+\rhoe/\rhoi}~
\end{equation}    
   defines the kink speed. 
Likewise, Equation~\ref{eq_TB_omgI} yields that
\begin{equation}
   \label{eq_TB_omgI_largeky}
\displaystyle 
  \frac{\omgI}{\omgR}
= -\frac{\pi}{8}
   \left(k_y l\right) g
   \frac{\rhoi-\rhoe}{\rhoi+\rhoe}~.
\end{equation}
To proceed, we note that $g$ now evaluates to 
   $2/\pi$, $1$, and $1/\sqrt{2}$ for the sine, linear, and parabolic
   profiles, respectively.
Consider the linear profile for now. 
Equation~\ref{eq_TB_omgI_largeky} recovers the cylindrical result
  obtained by \citet[][Equation~79b]{1992SoPh..138..233G}, 
  who were the first to adopt the jump conditions to examine quasi-modes
  from the dissipative-eigenmode perspective to our knowledge.
As suggested therein, the comparison between the slab and cylindrical
  configurations is possible by interpreting $R$ as the cylinder radius
  and prescribing $k_y$ with $1/R$. 
Indeed, adopting this prescription in Equation~\ref{eq_TB_omgI_largeky}
  further recovers the cylindrical results for the sine 
  \citep[][Equation~72]{2002ApJ...577..475R} 
  and parabolic profiles \citep[][Equation~7]{2014ApJ...781..111S}. 
Now we focus on the slab configuration.
For a linear profile, 
  Equations~\ref{eq_TB_omgk} and \ref{eq_TB_omgI_largeky} 
  have been obtained in an extensive series of studies
   using the ideal quasi-mode approach
  \citep[\eg][to name only a few early studies]{1978ApJ...226..650I,1979ApJ...227..319W}.
Among the quasi-mode studies, to our knowledge, TW98 were
  the only one that offered explicit
  expressions for the eigenfrequencies without requiring $\Theta \approx 1$.
It is reassuring to see that Equation~15 therein is exactly reproduced
  by our Equations~\ref{eq_TB_omgR} and \ref{eq_TB_omgI},
  which are derived with the dissipative eigenmode approach.
Neglecting the gas pressure, one readily finds that
  Equation~15a in TW98 is identical to Equation~\ref{eq_TB_omgR} here.
However, our Equation~\ref{eq_TB_omgI} with $g=1$ seems different 
      from Equation~15b in TW98.
This discrepancy is only apparent, and can be reconciled 
      by noting that $\Theta/(1+\Theta) = [1-\exp(-2 k_y R)]/2$.     
Our Equation~\ref{eq_TB_omgI_tmp} therefore generalizes the TW98 results
  by allowing the TL profile to be as arbitrary as the TB approach allows.
Likewise, it generalizes those dissipative eigenmode results where
  $\Theta \approx 1$ is assumed
  (Equation~57 in \citealp{1995JPlPh..54..129R},
  see also the references therein, in particular
  \citealp{1985JPlPh..33..199M};
  \citealp{1988JGR....93.5423H}).

To sum up at this point, the DR (Equation~\ref{eq_DR_TB}) 
   applies to rather arbitrary density profiles, provided $l/R \ll 1$.   
Furthermore, it is valid
   for arbitrary values of $k_y$ and $k_z$.
When $k_y^2 \gg |\omega^2/\va^2|$ and $k_y l \ll 1$, 
   the real part of the eigenfrequency
   is expressible by Equation~\ref{eq_TB_omgR}.
Likewise, the imaginary part is given 
   by Equation~\ref{eq_TB_omgI_tmp} for general profiles, and
   specializes to Equation~\ref{eq_TB_omgI} 
   for our profile choices.
However, in the more general situation,    
   Equation~\ref{eq_DR_TB} needs to be solved numerically
   after one specifies the parameters in the parantheses
   in Equation~\ref{eq_omega_formal}.
We adopt an iterative approach for this purpose, 
   starting with an initial guess for $\omgR$ such that
   a temporary value for $x_{\rm A}$ can be found.
We then evaluate the $\rho_{\rm A}/\rho'_{\rm A}$ term and
   solve Equation~\ref{eq_DR_TB} to yield $\omega$.
Adopting this updated $\omega$ as a further guess,
   we iterate this process until it converges to
   a unique eigenfrequency. 

\section{Numerical Results}
\label{sec_num_results}
We are now in a position to examine the dispersive properties
   of resonantly damped kink modes in coronal slabs, for which purpose
   it is admittedly difficult to exhaust
   the full parameter space as contained 
   in Equation~\ref{eq_omega_formal}. 
We choose to present our results in two parts.
Section~\ref{sec_sub_TBvalidation} will examine
   how the TB expectations compare with
   our self-consistent resistive results.
On the other hand, section~\ref{sec_sub_resistive} will 
   compare different profile specifications regarding their influence
   on the eigenfrequencies found with the resistive eigenmode approach.

\subsection{Comparison of Resistive Computations with Thin-Boundary Expectations}
\label{sec_sub_TBvalidation}

\begin{figure}
\centering
 \includegraphics[width=.8\columnwidth]{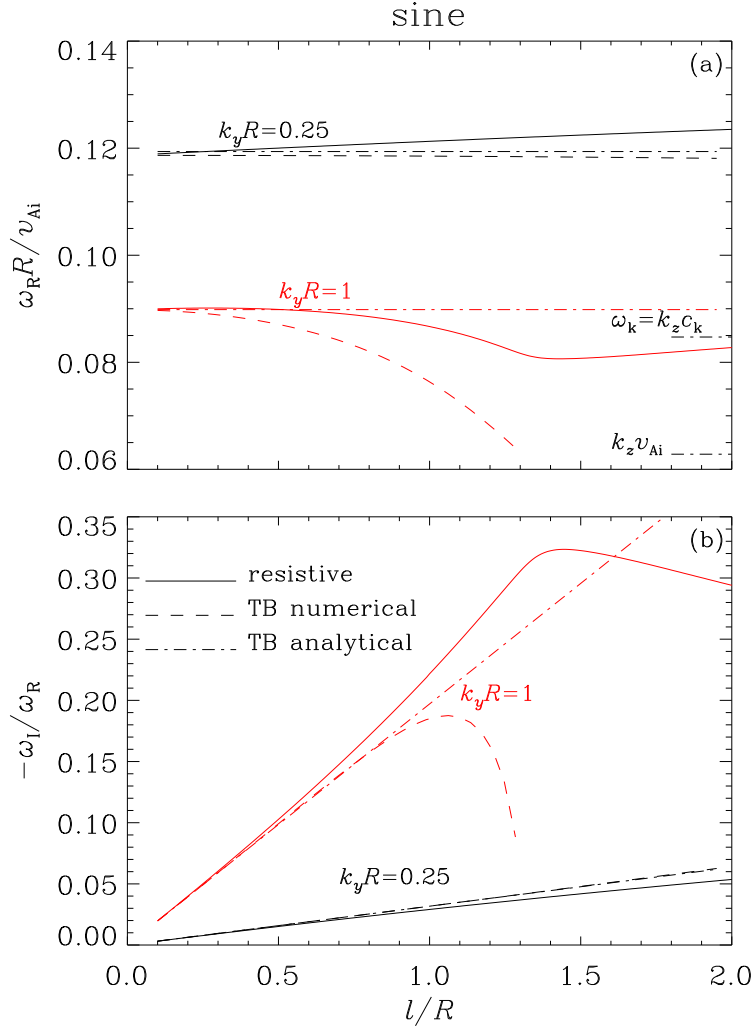}
 \caption{
 Dispersive properties of resonantly damped kink modes in coronal slabs.
 Shown here are the dependencies on the dimensionless TL width [$l/R$] of
         (a) the real part of the eigenfrequency [$\omgR$],
     and (b) the ratio of the imaginary to the real part [$-\omgI/\omgR$].
 The solid lines represent the results obtained with resistive computations,
     while the dashed ones correspond to the numerical solutions to the 
     explicit DR (Equation~\ref{eq_DR_TB}) valid in the thin-boundary (TB) limit
      ($l/R \ll 1$).
 The analytical solutions to this DR, nominally valid when 
     $k_y^2 \gg |\omega^2/\va^2|$ and $k_y l \ll 1$, are plotted
     by the dash--dotted curves
     (see Equations~\ref{eq_TB_omgR} and \ref{eq_TB_omgI} for details).
 Two different values of $k_y$ are examined as labeled.
 For comparison, the horizontal bars represent the angular frequencies $k_z \vai$ and 
     $\omega_{\rm k} = k_z c_{\rm k}$.
 A ``sine'' profile is adopted for the equilibrium density,
     the density contrast is fixed at $\rhoi/\rhoe = 10$,
     and the axial wavenumber is fixed at $k_z R = \pi/50$.  
 }
 \label{fig_omg_l_2ky}
\end{figure}

Our examination starts with Figure~\ref{fig_omg_l_2ky},
   where the dependencies on the dimensionless TL width ($l/R$)
   are shown for both 
   the real part of the eigenfrequency 
       ($\omgR$, Figure~\ref{fig_omg_l_2ky}a),
   and the ratio of the imaginary to the real part 
       ($-\omgI/\omgR$, Figure~\ref{fig_omg_l_2ky}b).
The equilibrium density profile is chosen to be 
   a ``sine'' one for illustration purposes. 
Furthermore, the density contrast and the axial wavenumber
   are fixed at $\rhoi/\rhoe = 10$ and $k_z R = \pi/50$, respectively. 
Two rather different values of $k_y$ are examined as represented 
   by the curves in different colors.       
The results from the self-consistent resistive computations
   are shown by the solid lines.
We also solve the explicit DR (Equation~\ref{eq_DR_TB})
   in the thin-boundary (TB) limit with both the iterative
   and analytical approaches, and label
   the corresponding results 
   with ``TB numerical'' (the dashed curves)
   and ``TB analytical'' (dash--dotted), respectively.
Two additional frequencies,  $k_z \vai$ and $\omega_{\rm k} = k_z c_{\rm k}$,
   are also plotted for comparison
   (see the horizontal ticks in Figure~\ref{fig_omg_l_2ky}a).
Examining the TB curves, one sees that the ``TB analytical" ones
   as given by Equations~\ref{eq_TB_omgR} and \ref{eq_TB_omgI}
   tend to satisfactorily reproduce the ``TB numerical'' solutions
   when $k_y l$ is small, which is evidenced by the close agreement
   between the black dashed and dash--dotted curves pertaining 
   to $k_y R = 0.25$. 
This is expected given that $k_y l \ll 1$ is assumed for deriving these expressions
   and that the additional assumption $k_y^2 \gg |\omega^2/\va^2|$ 
   tends to hold here. 
From Figure~\ref{fig_omg_l_2ky}a one also sees that the dash--dotted curves 
   may differ substantially from the kink frequency $\omega_{\rm k}$ when $k_y R$
   is not sufficiently large, suggesting the need to incorporate the $k_y$-dependence
   in analytical expressions for the eigenfrequency.
These details aside, it is more important to note that, relative to
   the ``TB analytical'' curves, the ``TB numerical'' ones tend to deviate more significantly from the resistive results. 
This happens even though the iterative solution to DR (Equation~\ref{eq_DR_TB})
   seems more self-consistent than its analytical counterpart. 
On top of that, the iterative approach does not converge when $l/R \gtrsim 1.3$ 
   for $k_y R = 1$, the reason being that $\omgR$ decreases towards $k_z \vai$
   as $l/R$ increases and eventually leaves the \Alf\ continuum.     
While only a ``sine'' profile is shown, we find that the resistive results tend
   to agree better with the ``TB analytical'' rather than the ``TB numerical'' results
   for other profiles as well.
We therefore conclude that, technically speaking, if one would like to invoke
   the TB limit to validate
   the eigenfrequencies numerically found for resonantly damped kink modes
   in coronal slabs with an equilibrium configuration similar to ours,
   then it is not worthwhile to further solve the relevant DR numerically.
Analytical expressions similar to Equations~\ref{eq_TB_omgR} and 
   \ref{eq_TB_omgI_tmp} serve this purpose better.        
In what remains, we use $\omega^{\rm TB}$ to denote the
   eigenfrequencies analytically expected with Equations~\ref{eq_TB_omgR} and 
      \ref{eq_TB_omgI}.
Likewise, by $\omega$ we mean the eigenfrequencies found from resistive computations.

How does $\omega^{\rm TB}$ compare with $\omega$ if we survey a broader set of parameters?
The following definitions are necessary,
\begin{equation}
  \delta \omgR 
= \frac{2\pi/\omgR^{\rm TB}}{2\pi/\omgR}-1,~~~
  \delta(\omgR/\omgI)
= \frac{\omgR^{\rm TB}/\omgI^{\rm TB}}{\omgR/\omgI}-1.
\label{eq_dOmg} 
\end{equation}
Evidently, $\delta\omgR$ and $\delta (\omgR/\omgI)$ defined this way pertain 
   to the wave period $P = 2\pi/\omgR$ 
   and damping-time-to-period ratio $\tau/P = |1/\omgI|/(2\pi/\omgR)$,
   respectively.
As such, they essentially comply with the definitions in the cylindrical study 
   by \citet[][Equation~(10)]{2014ApJ...781..111S}.
The only difference is that we evaluate the relative differences in
   the algebraic rather than the absolute sense.
Negative values of $\delta\omgR$ or $\delta(\omgR/\omgI)$ therefore represent where
   the TB expectations underestimate the true values of the relevant quantity.
In the latter case, this is equivalent to where
   the TB expectations overestimate the damping rates [$-\omgI/\omgR$].

\begin{figure}
\centering
 \includegraphics[width=1.\columnwidth]{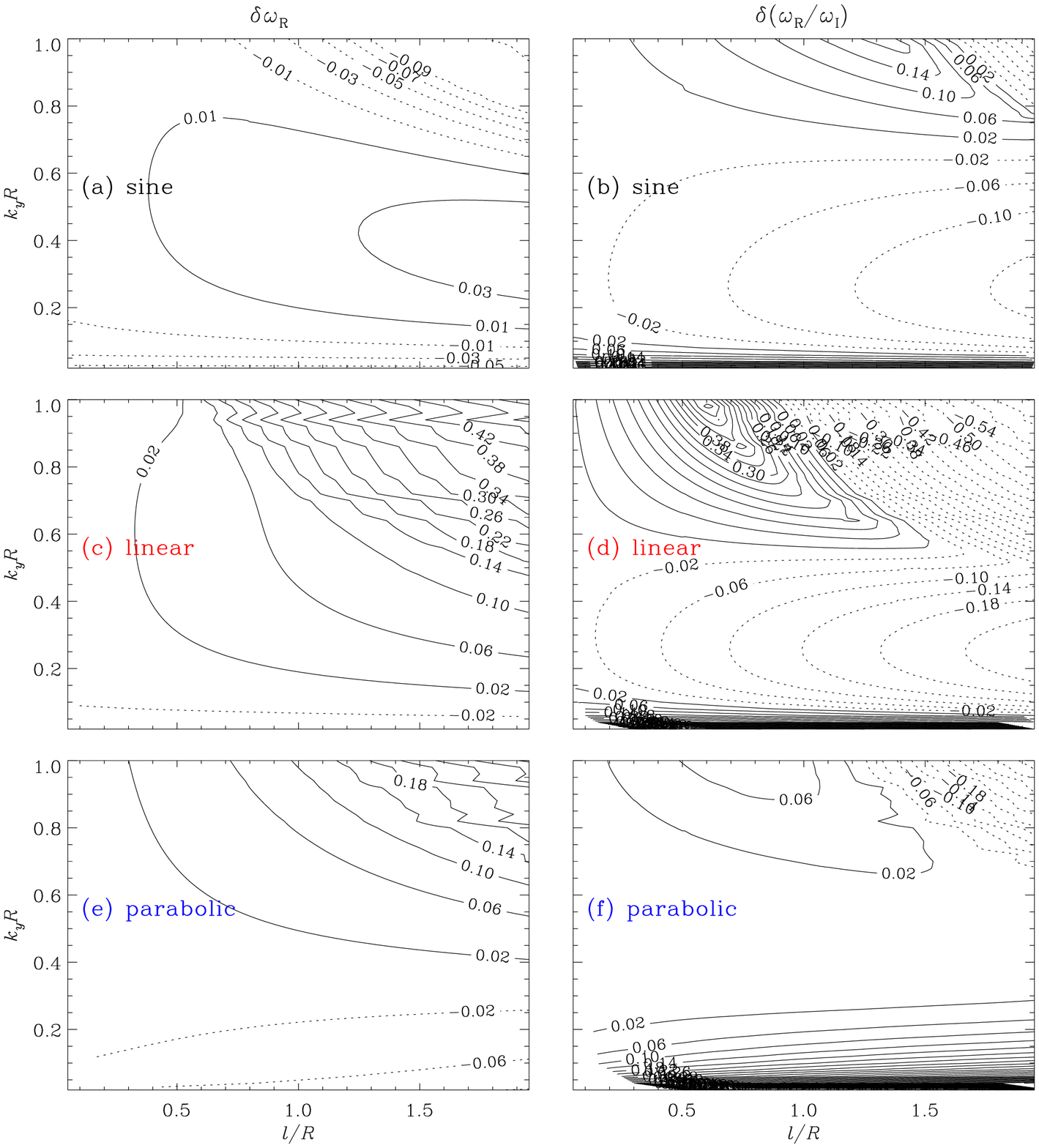}
 \caption{
Comparison of the eigen-frequencies self-consistently computed in resistive MHD
   with the results analytically expected in the thin-boundary limit. 
Shown here are the distributions in the $l/R-k_y R$ plane of
   $\delta \omgR$ (the left column) and $\delta(\omgR/\omgI)$ (right),
   the definitions for which are given in Equation~\ref{eq_dOmg}.
Different rows pertain to different density profiles are labeled. 
The contours in each panel are equally spaced, with negative (positive) values represented
   by the dotted (solid) curves. 
The pair of $[\rhoi/\rhoe, k_zR]$
   is fixed at $[10, \pi/50]$. 
 }
 \label{fig_resis_vsTB_varyKY}
\end{figure}

Figure~\ref{fig_resis_vsTB_varyKY} compares the TB values with
   the resistive results by showing $\delta\omgR$ (the left column)
   and $\delta(\omgR/\omgI)$ (right) as functions of the dimensionless TL width $l/R$
   and the out-of-plane wavenumber $k_y$.
Here a pair of $[\rhoi/\rhoe, k_z R] = [10, \pi/50]$ is adopted for
   the density contrast and axial wavenumber.
Furthermore, $k_yR$ ranges between $0.02$ and $1$,
   and $l/R$ ranges from $0.05$ to $1.95$.
Different profiles are examined in different rows as labeled. 
The contours in all plots are equally spaced, with negative (positive) values
   represented by the dotted (solid) curves.     
One sees that both $\delta\omgR$ 
   and $\delta(\omgR/\omgI)$ show some complicated dependencies on 
   $l/R$ and $k_y R$.
Let us first examine the lowermost portions where
   $k_y R$ is small such that the requirement $k_y^2 \gg |\omega^2/\va^2|$
   is not met for one to arrive at Equations~\ref{eq_TB_omgR} and 
   \ref{eq_TB_omgI}. 
One sees from the right column that $\delta(\omgR/\omgI)$ is consistently positive.
While the relevant contours are too crowded to read, a careful inspection
   indicates that $|\delta(\omgR/\omgI)|$ reaches
   $60\%$, $110\%$, and $240\%$ for the 
   sine, linear, and parabolic profiles, respectively. 
Nonetheless, that $k_y l$ is small in this portion makes it uninteresting in terms of
   wave damping.
More interesting is that the consistently negative $\delta \omgR$ reaches 
   merely $-5.6\%$, $-5.7\%$, and $-10.4\%$ for the sine, linear, and parabolic profiles,
   respectively.

Now move on to the portions where $k_y \gtrsim k_z$. 
Inspecting the left column, one sees that $\delta\omgR$ tends to be positive
    and increase with $l/R$ or $k_y R$ for the linear and parabolic profiles,
    peaking at $54\%$ ($36\%$) for the former (latter).
The pattern for the sine profile is more complicated, with
    a positive (negative) maximum of $4.2\%$ ($-10.1\%$)
    at $[l/R, k_yR] \approx [2, 0.38]$ (close to the upper-right corner).
Some profile dependence can also be told for $\delta(\omgR/\omgI)$ in the right column.
Common to all profiles is that $\delta(\omgR/\omgI)$ tends to posses a negative maximum
    at the upper-right corner, reading $-26\%$, $-68\%$, and $-43\%$
    for the sine, linear, and parabolic profiles, respectively.
Likewise, a positive maximum is attained at $k_yR \approx 1$, 
        reading $20\%$, $54\%$, and $8.5\%$ 
    when $l/R \approx 1.39$, $0.61$  and $0.82$ for the respective profiles.
However, a negative maximum exists towards $l/R \approx 2$ for the sine
    and linear profiles, 
    reaching $-15\%$ and $-25\%$ when 
     $k_yR \approx 0.26$ and $0.28$.
In contrast, for the parabolic profile, 
    the TB expectations show little difference from the resistive results
    in terms of $\omgR/\omgI$ regardless of $l/R$ 
    when $k_y R$ varies between, say, $0.2$ and $0.7$.

\begin{figure}
\centering
 \includegraphics[width=1.\columnwidth]{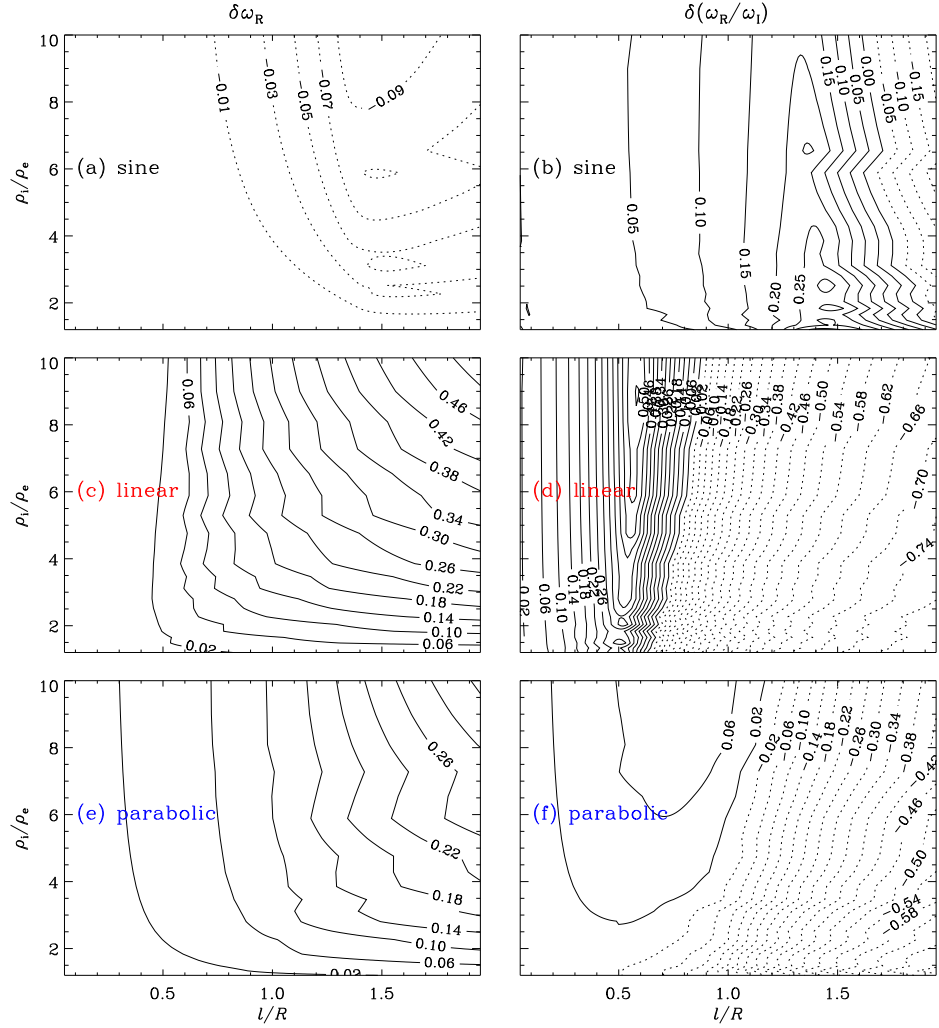}
 \caption{
Similar to Figure~\ref{fig_resis_vsTB_varyKY} but for 
   the distributions in the $l/R-\rhoi/\rhoe$ plane of
   $\delta \omgR$ (the left column) and $\delta(\omgR/\omgI)$ (right).
The pair of $[k_y R, k_z R]$
   is fixed at $[1, \pi/50]$. 
 }
 \label{fig_resis_vsTB_varyrhoie}
\end{figure}

Figure~\ref{fig_resis_vsTB_varyrhoie} presents, 
    in a format identical to Figure~\ref{fig_resis_vsTB_varyKY},
    how $\delta\omgR$ and $\delta(\omgR/\omgI)$ depend on
    $l/R$ and $\rhoi/\rhoe$ for a fixed pair of
    $[k_y R, k_z R] = [1, \pi/50]$.
The density contrast $\rhoi/\rhoe$ varies between $1.2$ 
    and $10$, a range that encompasses values representative
    of both active-region loops
       \citep[\eg][]{2004ApJ...600..458A}
       and streamer stalks \citep[\eg][Figure~2]{2011ApJ...728..147C}.
Furthermore, the dimensionless TL width $l/R$ varies from $0.05$
    to $1.95$. 
Examining the left column first, 
    one sees that overall $|\delta\omgR|$ tends to attain its maximum 
    when $\rhoi/\rhoe$ takes the largest value.
For the sine profile, $\delta\omgR$ tends to be negative, peaking at 
    $-10.2\%$ when $l \approx 1.43$. 
However, $\delta\omgR$ for the rest two profiles tends to be positive, 
    and attains $54\%$ ($36\%$) for the sine (parabolic) one around
    the upper-right corner. 
Now move on to the right column.    
For all profiles, one sees that $\delta(\omgR/\omgI)$ for a fixed $\rhoi/\rhoe$
    tends to switch from positive to negative values with increasing $l/R$. 
The negative maximum for the sine profile is attained
    at the upper-right corner, the specific value being $-26\%$. 
However, the corresponding maxima for the linear and parabolic profiles
    are both attained at the lower-right corner, reading 
    $-84\%$ and $-71\%$.
On the other hand, $\delta(\omgR/\omgI)$ for the sine profile attains a positive
    maximum of $32\%$ at $[l/R, \rhoi/\rhoe] \approx [1.4, 2.5]$.
For comparison, the positive maximum for $\delta(\omgR/\omgI)$ 
    attains $55\%$ ($8.5\%$) when $[l/R, \rhoi/\rhoe]$ reads roughly
    $[0.58, 9]$ ($[0.82, 10]$) for the linear (parabolic) profile.

It proves informative to compare our Figure~\ref{fig_resis_vsTB_varyrhoie} 
   with its cylindrical counterpart given by Figure~1 in \citet[][hereafter S14]{2014ApJ...781..111S},
   thereby highlighting some geometry-related differences
   in the behavior of resonantly damped kink modes.
Two remarks are necessary before proceeding, however.
Firstly, as suggested by \citet{1992SoPh..138..233G} and \citet{2008IAUS..247..228G},
   what we compute for kink modes in a slab geometry can be meaningfully compared
   with cylindrical results only when $k_y R$ takes integer values as far as the eigenfrequencies are concerned.
In particular, if $k_y R$ is specified to be unity, 
   then its cylindrical counterpart is the kink mode. 
Likewise, integer values of $k_y R$ larger than unity correspond to 
   fluting modes in the cylindrical geometry.
Secondly, we note that S14 adopted a value of $k_z R = \pi/100$ for
   the axial wavenumber, which is slightly different from what we specified.
We believe this slight difference is not important for our purpose, because both values of
   $k_z R$ pertain to the regime where the axial wavelength is much longer
   than the width of the hosting structure. 
Nonetheless, we avoid direct comparisons of numeric values in the two geometries.

With the necessary remarks, this comparison can be most readily presented on the basis
   of individual profiles.
For the sine one, the first row in Figure~\ref{fig_resis_vsTB_varyrhoie} 
   is morphologically different from its cylindrical counterpart (the first row
   in Figure~1 of S14), where $|\delta\omgR|$ tends to peak at 
   $[l/R, \rhoi/\rhoe] \approx [2, 6]$ and $|\delta(\omgR/\omgI)|$ tends to maximize
   either when $l/R \approx 1.1$ for large $\rhoi/\rhoe$ or 
          when $l/R \approx 2$   for small $\rhoi/\rhoe$.
Despite this, our slab computations agree with S14 in that the sine profile
   possesses the least difference between the TB expectations
   and self-consistently computed eigenfrequencies among the examined profiles.
For the linear profile, the second row in Figure~\ref{fig_resis_vsTB_varyrhoie} 
   is strikingly similar to the second row in Figure~1 of S14.
By this we mean that the maxima in $|\delta\omgR|$ and $|\delta(\omgR/\omgI)|$ 
   occur at very similar locations in the $l/R - \rhoi/\rhoe$ plane.
For the parabolic profile, a comparison between Figure~\ref{fig_resis_vsTB_varyrhoie} 
   and Figure~1 of S14 yields that the dependence of $|\delta\omgR|$ 
   on $l/R$ and $\rhoi/\rhoe$ is remarkably similar in both geometries.
However, there is an interesting difference in the behavior of $\delta(\omgR/\omgI)$,
    even though the locations where $|\delta(\omgR/\omgI)|$ takes its largest value
    are similar in the two geometries.
For this purpose we focus on the situation where $\rhoi/\rhoe \gtrsim 3$.   
While $|\delta(\omgR/\omgI)|$ is consistently less than $10\%$ for
   $l/R \lesssim 1$ in the slab case,
   some rather significant value for $|\delta(\omgR/\omgI)|$
   arises when $l/R$ exceeds $0.5$ in the cylindrical geometry.
In physical terms, this means that the TB expectation starts to overestimate
   the damping rates at some $l/R$ that is substantially larger in the slab 
   than in the cylindrical geometry.
We will return to this point later.   

\subsection{Effects of Profile Specifications on Eigenfrequencies in Resistive Computations}
\label{sec_sub_resistive}

This section focuses on the profile dependence of the eigenfrequencies of resonantly
    damped kink modes in coronal slabs.
For this purpose, we will present only the resistive results. 
In addition, the density contrast $\rhoi/\rhoe$ is fixed at $10$.

\begin{figure}
\centering
 \includegraphics[width=.9\columnwidth]{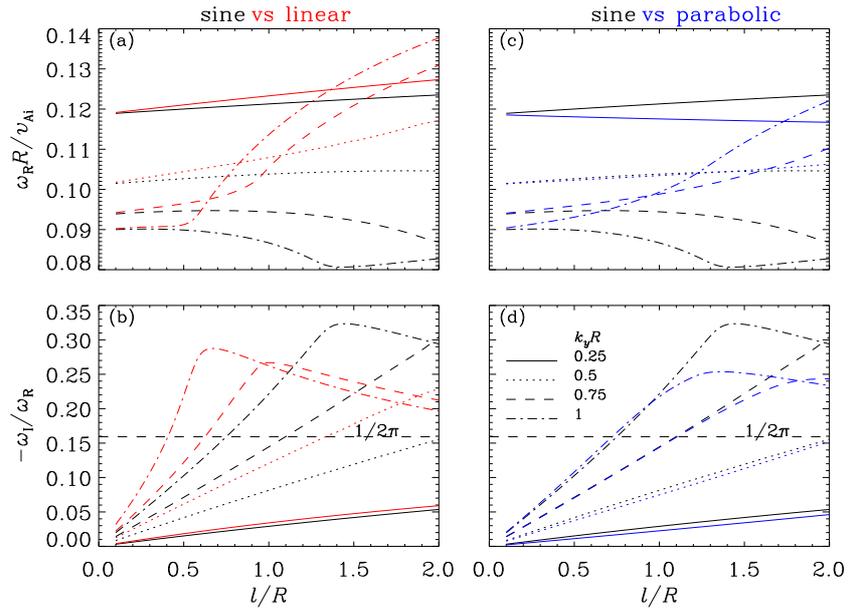}
 \caption{
Influence of the density profile prescription
   on the dispersive properties of resonantly damped kink modes in coronal slabs.
Shown here are the dependencies on the dimensionless TL width ($l/R$)
   of both the real part of the eigenfrequency ($\omgR$, the top row)
   and the ratio of the imaginary to the real part ($-\omgI/\omgR$, lower). 
Different profiles are differentiated by different colors, 
   and a number of $k_y$ are examined as represented by different linestyles. 
The horizontal dashed line in the lower row corresponds to $1/2\pi$, at which value
   the damping-time-to-period ratio attains unity.   
The pair of $[\rhoi/\rhoe, k_zR]$
   is fixed at $[10, \pi/50]$. 
 }
 \label{fig_omg_l_4ky_3prof}
\end{figure}

We start with Figure~\ref{fig_omg_l_4ky_3prof} 
   where the $l/R$-dependencies 
   of $\omgR$ (the upper row) and $-\omgI/\omgR$ (lower) are compared between
   the sine and linear (parabolic) profiles in 
   the left (right) column. 
Different colors are adopted to differentiate different profiles.
Likewise, different linestyles pertain to a number of different values
   for $k_y$.
The combination $[\rhoi/\rhoe, k_zR]$ is fixed at
   $[10, \pi/50]$.    
Overall, the point to draw is that 
   the profile specifications are important 
   for determining the eigenfrequencies in general, 
   and the damping rates in particular. 
Take the first row for instance.
For the examined values of $k_y$, Figure~\ref{fig_omg_l_4ky_3prof}a indicates that    
   the differences in $\omgR$ between the sine and linear profiles
   tend to increase with $k_y$, which is particularly true for thick TLs.
When $k_y R = 1$, the two profiles can hardly be distinguished in terms of $\omgR$
   for $l/R \lesssim 0.5$.
At larger $l/R$, however, $\omgR$ for the sine profile gradually decreases 
   to some minimum at $l/R \approx 1.4$ and then increases.
In contrast, $\omgR$ for the linear profile increases monotonically, 
   eventually exceeding its sine counterpart by $66\%$ when $l/R \approx 2$.
Moving on to Figure~\ref{fig_omg_l_4ky_3prof}b, one sees that
   $\omgR$ for the parabolic profile tends to be smaller than for the sine one 
   when $k_y R \lesssim 0.5$, with the tendency reversed for larger values
   of $k_y R$.
When $k_y R = 1$, the values of $\omgR$ may differ substantially between the two profiles,
   with the parabolic one exceeding the sine one by $48\%$ for $l/R \approx 2$. 
Now let us examine the second row.
A comparison of Figure~\ref{fig_omg_l_4ky_3prof}c with \ref{fig_omg_l_4ky_3prof}d
   indicates that in general the differences in $-\omgI/\omgR$ between the sine and linear
   profiles are more significant than those between the sine and parabolic profiles.
For sufficiently small $l/R$, this behavior is understandable given the profile dependence of
   the factor $g$ in Equation~\ref{eq_TB_omgI}.
Focusing on the case where $k_y R = 1$, one readily finds that $g$ evaluates to 
   $0.64$ and $0.66$ for the sine and parabolic profiles, respectively.
These values are substantially different from unity, the value that $g$ takes up
   for the linear profile.
The dash--dotted curves in Figure~\ref{fig_omg_l_4ky_3prof}d indicate that
   the linear $l/R$-dependence expected in the TB limit
   persists until $l/R \lesssim 1$ for the sine and parabolic profiles.
However, $-\omgI/\omgR$ for the linear profile starts to deviate from the TB behavior
   at considerably smaller $l/R$.
In fact, the red dash--dotted curve in Figure~\ref{fig_omg_l_4ky_3prof}c suggests that 
   $-\omgI/\omgR$ reaches some maximum around $l/R \gtrsim 0.6$ and then decreases
   with $l/R$.  
As a result, the values that $-\omgI/\omgR$ attain at $l/R \gtrsim 0.6$
   for the linear profile may exceed those for the sine one by as much as $126\%$.

With Figure~\ref{fig_omg_l_4ky_3prof} we further remark on the similarities
    and differences in the eigenfrequencies of resonantly damped kink modes
    between the slab and cylindrical geometries.
We examine only the linear and parabolic profiles for this purpose, given the difficulty
    to exhaust the comparison between the geometries.
In addition, we focus on the case $k_y R = 1$ such that this comparison is meaningful.    
For the linear one, we note that the $l/R$-dependencies of $\omgR$ and $-\omgI/\omgR$
    are qualitatively similar to their cylindrical counterpart as given by 
    Figure~5 in \citet[][S13]{2013ApJ...777..158S}.
In particular, the values of $\omgR$ in both geometries
    nearly stay constant at small $l/R$ but then increase substantially with $l/R$.
Likewise, a nonmonotonical $l/R$-dependence of $-\omgI$ is common to both geometries.
While not shown, the plot of $\omgI$ rather than $-\omgI/\omgR$
    is remarkably similar to Figure~5 in S13.
    We also note that this behavior of the eigenfrequencies was originally found
    in the slab computations by TW98 to our knowledge.
It is worth noting that this remarkable similarity is found even though
    S13 employed a different combination $[\rhoi/\rhoe, k_z R]$ of $[5, \pi/100]$. 
The comparison of the parabolic results between the two geometries is interesting from
    another perspective,   
   for which purpose we note that the horizontal dashed lines in 
   the lower row in Figure~\ref{fig_omg_l_4ky_3prof} represent $1/2\pi$.
Evidently, this corresponds to the situation where the ratio of
   the damping time to period reaches unity, which may serve as
   a dividing line separating the under-damped from
   over-damped modes. 
As implied by Figure~5 in S13 and further corroborated by Figure~1 in S14,
   kink modes in the cylindrical geometry can hardly become over-damped.  
This happens because while the TB expressions readily
   predict over-damping for sufficiently large $l/R$,
   the actual values of $-\omgI/\omgR$ at those values of $l/R$ are 
   smaller than the TB expectations.
In the slab geometry, however, the $l/R$-dependence of $-\omgI/\omgR$
   starts to deviate from the TB expectations only when $l/R \gtrsim 1$.
For comparison, Equation~\ref{eq_TB_omgI} predicts that $-\omgI/\omgR$ 
   exceeds $1/2\pi$ when $l/R \ge 0.78$.
As a result, the blue dash--dotted curve in Figure~\ref{fig_omg_l_4ky_3prof}d
   stays above the horizontal line for $l/R \gtrsim 0.72$, and peaks at
   $0.25$ despite that the curve decreases with $l/R$ eventually.
We note by passing that fluting modes in the cylindrical geometry can readily
   become over-damped by resonant absorption in the \Alf\ continuum
   \citep{2017ApJ...850..114S}.     

\begin{figure}
\centering
 \includegraphics[width=.8\columnwidth]{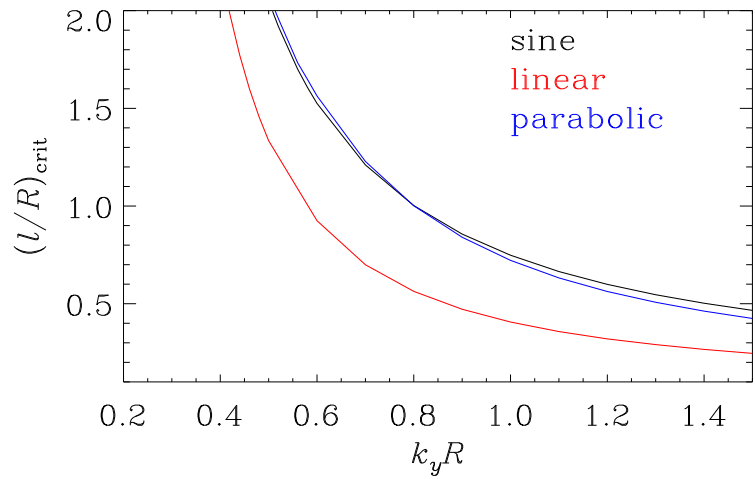}
 \caption{
Influence of the density profile prescription
   on the dispersive properties of resonantly damped kink modes in coronal slabs.
Shown here is the dependence on $k_y$ of the 
   critical dimensionless TL width $(l/R)_{\rm crit}$, at which value
   $|\omgI/\omgR|$ attains $1/2\pi$.
Different colors are adopted to represent different density profiles. 
The pair of $[\rhoi/\rhoe, k_zR]$
   is fixed at $[10, \pi/50]$. 
 }
 \label{fig_lrCrit_ky_3prof}
\end{figure}

Distinguishing the over-damped from under-damped situations 
   can also help bring out the profile dependence.
Looking at the lower row in Figure~\ref{fig_omg_l_4ky_3prof}
   and specializing to a profile, 
   at a given $k_y$ we can deduce a critical TL width $(l/R)_{\rm crit}$ beyond which
   $-\omgI/\omgR$ starts to exceed $1/2\pi$.
Surveying a more extensive range of $k_y$ then leads to
   Figure~\ref{fig_lrCrit_ky_3prof}, where $(l/R)_{\rm crit}$
   is plotted as a function of $k_y$ in different colors for different profile
   choices.     
One sees that the curves labeled ``sine'' and ``parabolic'' are close to one another,
   understandable given the insignificant differences in $-\omgI/\omgR$ 
   between the two profiles
   at sufficiently small $l/R$. 
However, these two curves deviate considerably from the curve labeled ``linear'',
   which is true throughout the entire range of $k_y$ examined here.      

\begin{figure}
\centering
 \includegraphics[width=.6\columnwidth]{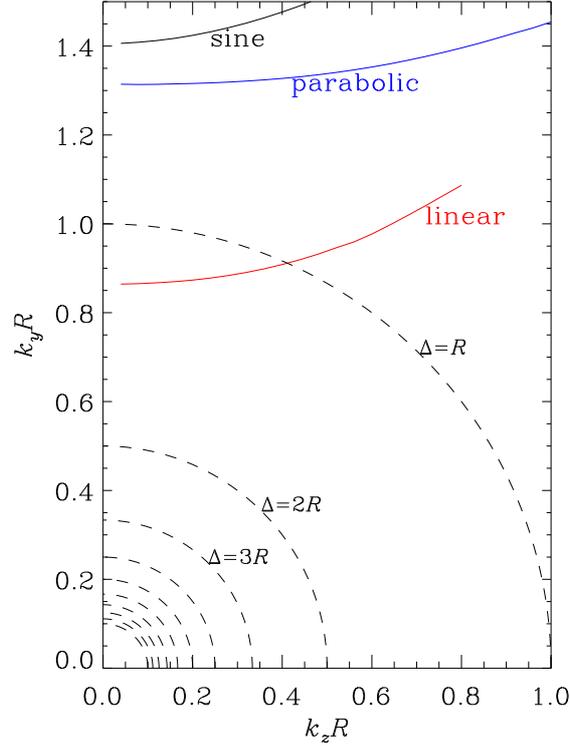}
 \caption{
Influence of the density profile prescription
   on the dispersive properties of resonantly damped kink modes in coronal slabs.
The solid curves in different colors pertain to the pair of $[k_y, k_z]$
   where $|\omgI/\omgR|$ attains $1/2\pi$ at a fixed $l/R=0.5$
   for different density profiles as labeled.
The density contrast is fixed at $\rhoi/\rhoe = 10$.
Any dashed curve corresponds to those combinations of $[k_y, k_z]$ where
   the power spectral density drops from its maximum by a factor of $e$
   for some initial perturbation that depends on $x$ and $y$ as
   $\exp[-(x^2+y^2)/2\Delta^2]$.
A uniform spacing of $R$ is adopted for $\Delta$ to plot the dashed curves,
   the outermost one pertaining to $\Delta = R$.
See the text for details.    
 }
 \label{fig_PSD}
\end{figure}

The computations so far correspond to a fixed axial wavenumber
   of $k_z R = \pi/50$, which is more relevant for standing kink modes trapped
   in, say, AR loops or AR arcades.
However, examining other values of $k_z$
   will be informative to say the least, and we look back at 
   the parameters in the parentheses in Equation~\ref{eq_omega_formal} for this purpose.
If we allow only $k_y$ to vary and construct a plot similar to 
   Figure~\ref{fig_omg_l_4ky_3prof}b, 
   then we end up with a critical $k_{y, {\rm crit}}$ that corresponds
   to $-\omgI/\omgR = 1/2\pi$. 
Further varying $k_z$ therefore leads to a series of $k_{y, {\rm crit}}$, 
   and the output from such a practice with $l/R = 0.5$ is plotted 
   in Figure~\ref{fig_PSD}.
Here different density profiles are represented by the solid curves in different colors. 
When placed in the $k_y$-$k_z$ plane, 
   the under-damped (over-damped) modes then lie below (above) 
   the relevant solid curve for a specific profile prescription.  
Above all, one sees once again the importance of profile prescriptions
    for determining the dispersive properties of resonantly damped modes.
When examined this way, not only does the curve labeled ``linear'' lie
    at quite some distance away from the rest, 
    but also some substantial difference arises 
    between the curves labeled ``sine'' and ``parabolic''.
    
Figure~\ref{fig_PSD} allows us to say a few words on 
    impulsively generated kink waves in coronal slabs, in which context
    streamer waves are likely to be relevant. 
Suppose that these waves result from
    the lateral impingement on a coronal slab by, say, a bulk of mass
    (coronal mass ejections in the case of streamer waves, see, \eg\
    \citealp{2010ApJ...714..644C};
    \citealp{2011SoPh..272..119F},
    \citealp{2020ApJ...893...78D}).     
It seems observationally difficult to tell the characteristics
    of the interaction between the slab and, say, the mass motion.
To make some quantitative progress, we nonetheless assume that the duration
    of the interaction is shorter than the 
    (quasi-)periods of the resulting waves. 
We further assume that the interaction is spatially isotropic 
    in the $x$-$y$ plane (see Figure~\ref{fig_EQprofile}a)
    and specifically possesses a Guassian form
        $\exp[-(x^2 + y^2)/2\Delta^2]$.
Evidently, $\Delta$ measures the spatial extent of the interaction.
The power spectral density (PSD) is then $\propto \exp[-(k_y^2 + k_z^2) \Delta^2]$, 
    and therefore decreases from its maximum by a factor of $e$
    along a circle of radius $1/\Delta$ in the $k_y$-$k_z$ plane.
The dashed circles in Figure~\ref{fig_PSD} correspond to
    a series of $\Delta$ uniformly spaced by $R$, 
    with the outermost one corresponding to $\Delta = R$. 
Two situations then arise regarding the damping
    of the components
    constituting an impulsive wave. 
If $\Delta \gtrsim 2~R$, then 
    the majority of the components can survive resonant damping.
If $\Delta \lesssim R$ instead, then the heavy resonant damping
    of the components pertaining to the upper-left corner 
    in Figure~\ref{fig_PSD} will make it difficult for
    them to propagate to large distances. 
While these statements are enabled by Figure~\ref{fig_PSD}, 
    this figure alone does not allow us to 
    tell the eventual temporal and spatial variations
    of impulsively generated waves.
To name but one intricacy, we note that wave dispersion is well known
    to play a critical role in the dynamics of impulsive waves in 
    both cylindrical \citep[\eg][]{1983Natur.305..688R,1986NASCP2449..347E,2014ApJ...789...48O,2015ApJ...806...56O,2015ApJ...814..135S,2016ApJ...833...51Y,2017ApJ...836....1Y} 
    and Cartesian equilibria
    \citep[\eg][]{1986GeoRL..13..373E,1993SoPh..144..101M,2004MNRAS.349..705N,2013A&A...560A..97P,2016ApJ...826...78Y,2018ApJ...855...53L,2019A&A...624L...4G}.
Of particular relevance is the one by
    \citet{1993SoPh..144..101M}, who performed two-dimensional (2D)
    time-dependent simulations to examine the response of a coronal slab
    to a lateral exciter.
An extension of this study to 3D will clarify the behavior of the impulsively generated
    waves when the spatial extent of the driver is finite in the $y$-direction.
Consequently, our eigenmode analysis involving a finite $k_y$ will be much relevant, 
    as we have practiced previously in other contexts
    \citep[\eg][]{2017ApJ...836....1Y, 2018ApJ...855...53L}.

\section{Summary}
\label{sec_conc}
This study was motivated by the apparent lack of an extensive survey 
   of the parameters that may influence the resonant damping 
   of kink modes in straight coronal slabs.
To this end, we worked in the framework of pressure-less, gravity-free, resistive 
   MHD and adopted an eigenvalue problem (EVP) approach. 
The equilibrium density was restricted to be structured only in the $x$-direction,
   but was allowed to take a rather generic form comprising 
   a uniform interior, a uniform exterior, and a continuous transition layer (TL)
   in between.     
Inspired by examinations on resonantly damped kink modes in coronal cylinders
   \citep{2013ApJ...777..158S,2014ApJ...781..111S},
   we paid special attention to the effects of the mathematical form
   for describing the density profile (``profile'' for brevity).
By doing this we generalized the study by \citet{2007SoPh..246..213A}, who addressed
   an identical problem but nonetheless took a particular profile
   and examined only the influence of the out-of-plane wavenumber $k_y$.
We additionally conducted an analytical study in the thin-boundary (TB) limit, 
   and the resulting expressions generalized previous results 
   by allowing a broader set of profile choices and/or 
   a broader range of $k_y$.   

Our results are summarized as follows: 
Technically speaking, we find that the eigenfrequencies computed with the
   self-consistent resistive approach tend to agree better with the analytical 
   rather than the numerical solutions to the
   dispersion relation (Equation~\ref{eq_DR_TB})
   derived in the TB limit.
We therefore recommend expressions similar to 
   Equations~\ref{eq_TB_omgR} and \ref{eq_TB_omgI_tmp}
   for one to validate numerical studies on resonantly damped kink modes
   in an equilibrium similar to ours.
Physically speaking, we find that the eigenfrequencies in general, 
   and the damping efficiency in particular, are sensitive to  
   how the density profile is formulated. 
In particular, the ratio of the imaginary to the real part of the eigenfrequency
   for the linear profile can readily exceed its sine counterpart
   by a factor of two or so. 
When quantifying the deviation of the self-consistently computed
   eigenfrequencies from the TB expectations,
   we highlighted some differences in resonantly damped kink modes
   between the slab and cylindrical geometries.
An exemplary result pertains to the parabolic profile, 
   for which we find that kink modes in the slab (cylindrical) case can
   readily (hardly) become over-damped.
We also connected our computations to
   impulsively generated kink waves in coronal slabs by assuming 
   that the impulsive driver depends on $x$-$y$ in a Gaussian form
   characterized by a spatial scale $\Delta$.
In the parameter range that we examined, we find that    
   resonant absorption becomes relevant when $\Delta \lesssim R$, 
   strongly damping those components with $k_y \gtrsim 1/R$
   where $R$ is the slab half-width. 

Coronal structures are admittedly far more complicated than modeled here,
   in which context we name only two intricacies before closing:
First, it proves observationally difficult to tell the mathematical form
   for the density profile, one primary culprit being that the corona
   is optically thin in, say, the EUV. 
Seeing coronal loops as cylinders, one way to tackle this issue has been to 
   experiment with a number of profiles and contrast the forward-modeled
   EUV intensities with measurements by, say, TRACE 
   \citep[\eg][]{2003ApJ...598.1375A} 
   or the Atmospheric Imaging Assembly
   on board the Solar Dynamics Observatory 
     \citep[SDO/AIA, \eg][]{2017A&A...600L...7P}. 
However, while there is strong evidence that the density distribution is
   continuous rather than discontinuous, a statistical survey
   indicates that it is difficult to further discriminate between
   different profiles \citep{2017A&A...605A..65G}.
On this aspect we note that evidence-based model comparison
   in the Bayesian framework is a promising way forward
   (\eg\ \citealp{2015ApJ...811..104A};
         \citealp{2021ApJS..252...11A},
   see also the review by
      \citealp{2018AdSpR..61..655A}).           
Second, structural curvature needs to be accounted for when oscillations in,
   say, active-region arcades are examined.
A step forward is to consider a cylindrical system $(r, \theta, y)$ 
   and assume that the equilibrium magnetic field
   $\myvect{B}_0$ is of the form $\hat{\theta}/r$.
When $k_y \ne 0$, resonant absorption was shown to remain relevant by 
   \citet{2018ApJ...858....6H}, who nonetheless assumed a particular 
   density profile
   to ensure that kink modes are trapped.
For other profiles, however, lateral leakage is in general
   operational \citep[\eg][]{2006A&A...446.1139V,2006A&A...455..709D}, 
   even though this effect tends to weaken with $k_y$ 
   \citep{2013ApJ...763...16R}. 
We take these intricacies as strengthening rather than weakening the need
   for further examinations of coronal kink modes in a slab configuration.
Regarding the first intricacy, that the eigenfrequencies are sensitive to 
   profile choices can be employed in conjunction with
   the forward-modeling approach to tighten the constraints
   on the profile description, 
   as has been practiced by \citet{2018ApJ...860...31P}, who adopted
   a cylindrical geometry. 
Likewise, the second intricacy means that a systematic study is needed to
   assess the importance of resonant absorption relative to lateral leakage for
   kink modes in curved slabs, which has been initiated by \citet{2013ApJ...763...16R}. 
All of these developments are warranted, but nonetheless are
   left for future studies.    

\begin{acks}
We thank the referee for his/her constructive comments, which helped improve
    the manuscript substantially.
This research was supported by the 
    National Natural Science Foundation of China
    (HY:41704165, BL: 41674172, 11761141002, 41974200; SXC:41604145)
    and by Shandong University via grant No 2017JQ07.
We also acknowledge the International Space Science Institute-Beijing
   (ISSI-BJ) for supporting the international teams
   ``MHD Seismology of the Solar Corona"
   and ``Pulsations in solar flares: matching observations and models".    
\end{acks}

\footnotesize{
\paragraph*{Disclosure of Potential Conflicts of Interest} The authors declare that they have no conflicts of interest.
}

\bibliographystyle{spr-mp-sola}
\bibliography{seis_generic}


\clearpage

\clearpage

\clearpage

\clearpage

\clearpage

\clearpage

\clearpage

\end{article}

\end{document}